\documentclass[manuscript,screen]{acmart}

\AtBeginDocument{%
  \providecommand\BibTeX{{%
    \normalfont B\kern-0.5em{\scshape i\kern-0.25em b}\kern-0.8em\TeX}}}

\acmConference[PodRecs '20]{PodRecs '20: ACM Conference on Recommender
  Systems}{September 22--26, 2020}{Online}

\begin{document}

\title{A review of metadata fields associated with podcast RSS feeds}

\author{Matthew Sharpe}
\email{msharpe@spotify.com}
\orcid{1234-5678-9012}
\affiliation{%
  \institution{Spotify}
  \streetaddress{Regeringsgatan 13}
  \city{Stockholm}
  \state{Stockholm}
  \postcode{112-21}
  \country{Sweden}
}


\renewcommand{\shortauthors}{Sharpe}

\begin{abstract}
Podcasts are traditionally shared through RSS feeds. As well as pointing to the audio files, RSS gives a creator a way of providing metadata about the podcast shows and episodes. We investigate how certain metadata fields associated with podcasts are currently being used and comment on their applicability to recommendations. Specifically, we find that many creators are not using the {\tt itunes:type} field in the expected fashion, and that using this field for recommendations might not lead to an optimal user experience. We perform similar explorations for the season number and the category associated with a podcast, and also find that the fields aren't being used in the expected fashion. Finally, we examine the notion that a single podcast show is the same as a single RSS feed. This also turns out to not be strictly true in all cases. In short, the metadata associated with many podcasts isn't always reflective of the show and should be used with caution.
\end{abstract}





\maketitle

\section{Introduction}
Around 155 million Americans have listened to a podcast, with the number of weekly podcast listeners growing from approximately 10 million in 2015, up to an estimated 24 million in 2020\cite{PodGrowth}. It's also estimated that the number of new podcast shows published grew from 58,115 in 2015 to over 307,000 in 2019.\cite{PopularCategories}. With a huge increase in the number of shows and the number of consumers, the problem of recommending podcasts to users becomes more important. Previous work has seen podcast recommendations generated by using information about the user's intention on signing up for a platform\cite{10.1145/3308558.3313540}, using a user's music tastes\cite{10.1145/3397271.3401101} and through non-textual characteristics of the podcasts (such as the seriousness and the energy)\cite{morethanwords}. Traditionally, podcasts have been distributed entirely through RSS feeds. An RSS (really simple syndication) feed is a dialect of XML that can be used to publish regularly updated content, like podcasts\cite{RSSDefinition}. These feeds primarily serve to point any client towards the audio file that the user might want to play. However, along with the raw directions to the audio file, the creator of the podcast can also provide metadata. This metadata can provide information at the show level or at the specific episode level. Previous research has used the presence and length of certain metadata fields as an input to recommendations\cite{podcred}, but hasn't directly used the values provided. In this paper, we focus specifically on elements of the show level metadata and on how they're currently being used. Podcast creators typically set their own metadata via their podcast hosting platform, and the metadata applies across any listening platform where the show can be heard. A full guide of the required and optional fields can be found online\cite{AppleGuide} and an example of an RSS feed can be found in the appendix.

Important metadata fields to note include:

\begin{itemize}
\item {{\tt title}}: The title of the podcast show.
\item{{\tt itunes:type}}: Sometimes referred to as the consumption order, as this is usually used to inform the client which order the podcast is best consumed in. Values include:
\begin{itemize}
    \item{{\tt episodic}}: Used for shows that should be consumed without any specific order (the default value - traditionally ordered from the newest episode to the oldest episode\cite{AppleGuide}.)
    \item{{\tt sequential}}: Used for shows that should be consumed in sequential order (ordered from the oldest episode to the newest episode\cite{AppleGuide}).
\end{itemize}
\item{{\tt itunes:season}}: Used to show the season number the episode belongs to.
\item{{\tt itunes:category}}: Used to specify the category the show belongs to, where common categories include\cite{PopularCategories}:
    \begin{itemize}
        \item Society \& Culture
        \item Education
        \item Religion \& Spirituality
    \end{itemize}
\end{itemize}

Each of these fields could be used to enhance or augment the user's podcast experience, whether through modifying the client to improve playback, or using the information to better recommend podcasts to the user. Here, we investigate how the metadata fields are currently being used and whether we believe they are appropriate to use without any processing.

\section{Consumption Order}
As noted in the introduction, one of the elements of many podcast RSS feeds is the {\tt itunes:type} field. Historically, it is one of the most-used metadata fields\cite{TagUsage}. As well as being used to specify the consumption order of the podcast, this field could also be used to infer the type of experience the user is likely to want with a podcast.

As an example, a large majority of news podcasts are episodic - that is, users are unlikely to want to listen to an episode that is already many days (or weeks or years) out of date. Therefore, in designing a user experience (whether through recommendations, search, marketing etc.) it would likely be a bad idea to heavily feature the first episode of an episodic podcast.

Conversely, more fiction and true crime shows are labelled as sequential. That is, they're designed to be played like most TV shows - starting at episode 1 (least recently published) and proceeding until the end (most recently published). Featuring anything other than the first episode for users who are new to the show is likely to be a fairly bad user experience.

Across Spotify's entire podcast catalogue, 97.5\% of all podcast shows are listed as being episodic. However, there are variations across the different categories, shown in tables 2 and 3.


\begin{table}
  \caption{Percentage of a category's shows that are listed as episodic, for the categories containing the highest percentage of episodic shows}
  \label{tab:freq}
  \begin{tabular}{ccl}
    \toprule
    Show Category&Percentage of shows that are episodic\\
    \midrule
    Sport&99.0\% \\
    Arts&98.2\% \\    
    News&98.2\% \\
    Business&98.0\% \\    
  \bottomrule
\end{tabular}
\end{table}

\begin{table}
  \caption{Percentage of a category's shows that are listed as serial, for the categories containing the highest percentage of serial shows}
  \label{tab:freq}
  \begin{tabular}{ccl}
    \toprule
    Show Category&Percentage of shows that are serial\\
    \midrule
    Fiction&11.7\% \\
    True Crime&9.0\% \\    
    Games \& Hobbies&6.5\% \\
    History&4.8\% \\
  \bottomrule
\end{tabular}
\end{table}

\subsection{Is the given show type accurate?}

It may be surprising to learn that only 11.7\% of fiction podcasts are designed to be listened to from oldest to newest. And that news podcasts are more likely to be designed to be consumed in this way (oldest to newest) than arts podcasts.

Looking at specific examples, the Office Ladies podcast is a popular podcast going through the American version of The Office - starting with the first episode and working its way through to the last episode\cite{OfficeLadies}. \textbf{From first to last} episode. And yet the {\tt itunes:type} of the podcast is {\verb|episodic|} - any client or recommendation algorithm using this field will suggest the user starts with the newest episode, not the oldest episode.

Revisionist History is a popular podcast revisiting historical events and reinterpreting them\cite{RevisionistHistory}. Each episode is a stand-alone and it doesn't require knowledge of a prior episode in order to understand the most recent episode. And yet, the {\tt itunes:type} of the podcast is {\verb|serial|} - such that the user will be encouraged to start with the oldest episode and work their way through to the newest episode.

While it's impossible for us to say at this point exactly how many {\tt itunes:type} run contrary to expectations, we can infer the size of the error by looking at what new users do when presented with the podcast. As we use the {\tt itunes:type} to help determine the order in which we present the episodes, we introduce a bias whereby we encourage a user to start with the episode suggested by the {\tt itunes:type} tag.

Consequently, if the majority of users choose to listen to the newest episode first, it's more likely that the podcast should be consumed in that way (newest to oldest) than if the majority of users choose to listen to the oldest episode first.

\begin{figure}[h]
  \centering
  \includegraphics[width=\linewidth]{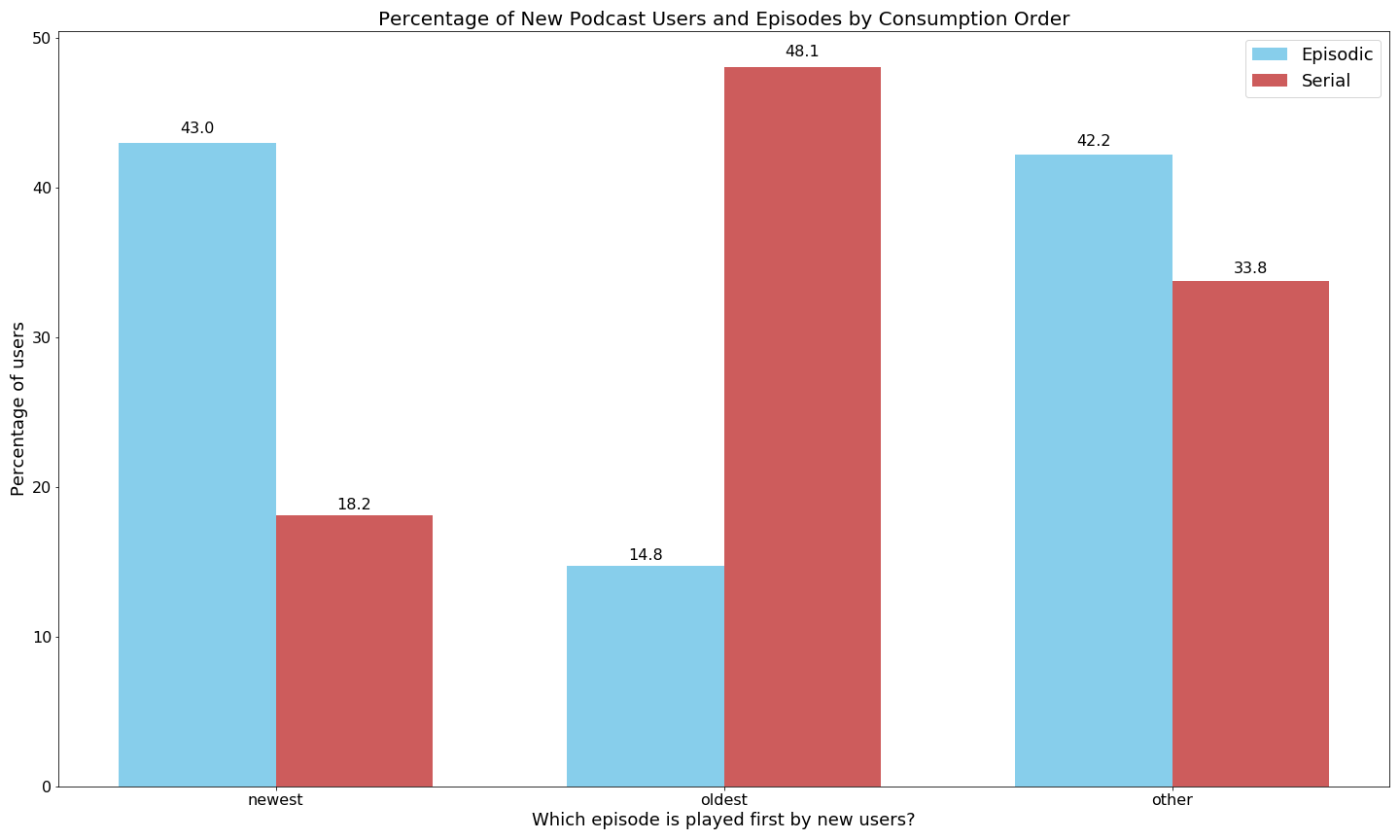}
  \caption{For podcast users who are new to a show (with five or more episodes) on Spotify in July 2020, which percentage of them listen to the two newest episodes, the two oldest episodes, or other episodes, first.}
  \Description{Clever commentary}
\end{figure}

As we see in figure 1, while the general order is as we would expect (the most common behaviour for new users on serial podcasts is to play one of the oldest two published episodes, the most common behaviour for new users on episodic podcast is to play one of the newest two published episodes), there are lots of users and podcasts for which this behaviour isn't seen\footnote{We use two episodes for both of these cases to account for the fact that many podcasts publish an initial teaser episode that many users choose to skip, and because of the phenomena detailed later on - that of multiple episode types.}. Many users listen to episodic podcasts starting at the oldest episode, and many users listen to serial podcasts starting at the newest episode.

There's also a large category of podcasts for which neither of the default orderings (newest to oldest, or oldest to newest) are necessarily the best ordering - shows included in the other category include many podcasts that feature stand-alone episodes. History podcasts that treat each episode as an independent story, for example.

In conclusion, though the {\tt itunes:type} field might give some indication of the type of podcast, in many cases we found the values present to be misleading and/or confusing. We advise proceeding with caution if using this field to help inform any podcast features (including recommendations).

\section{Season Number}

Much in the same way that television shows are often broken up into seasons, a podcast can be broken up into multiple seasons as well.

\begin{figure}[h]
  \centering
  \includegraphics[width=\linewidth]{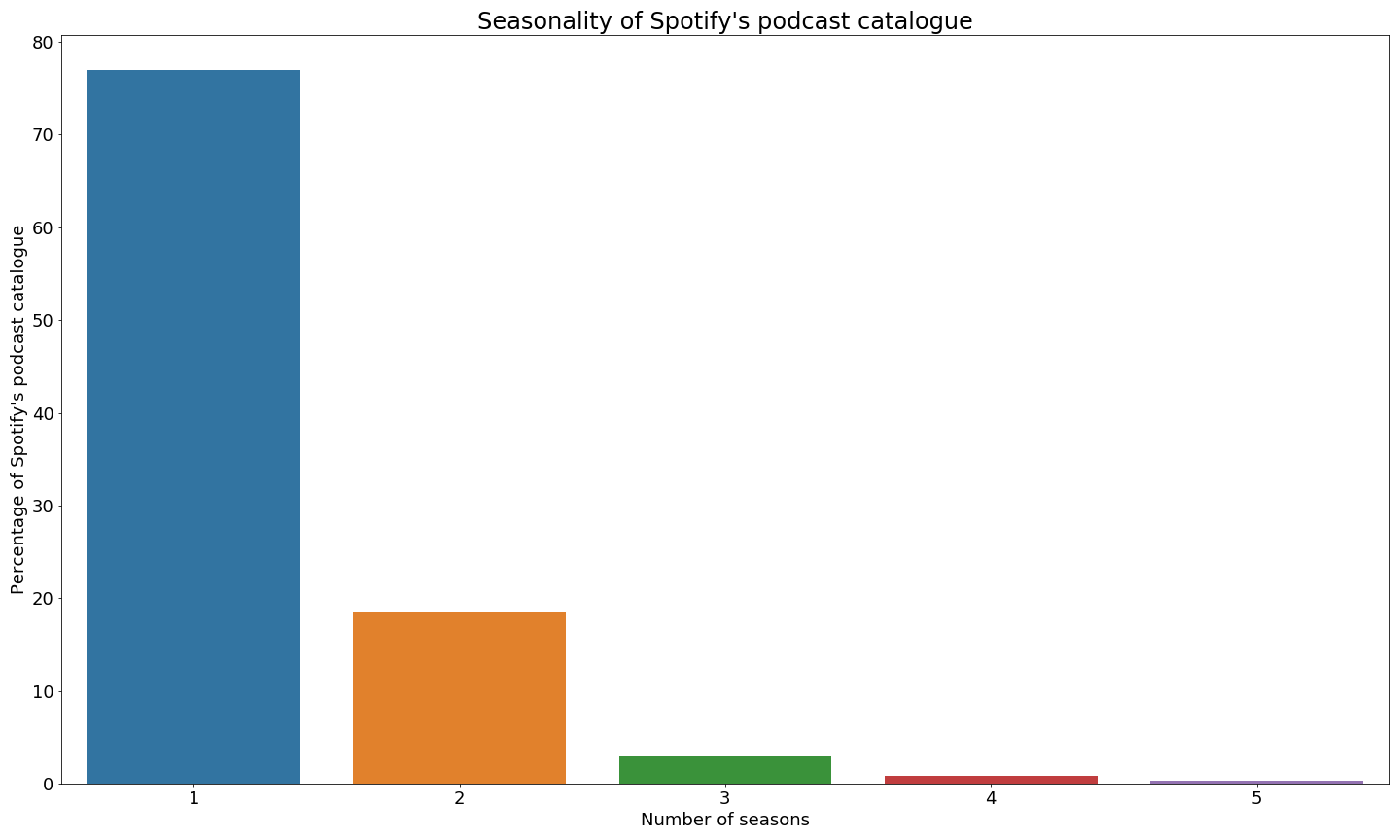}
  \caption{Number of podcast seasons per podcast show on Spotify}
  \Description{A graph showing the distribution of the number of seasons that podcas shows on Spotify have.}
\end{figure}

Figure 2 shows that the large majority of podcasts aren't seasonal. Around 99.6\% of all podcasts on Spotify have five or fewer seasons, with over 75\% of podcasts having one season only (and hence not being seasonal).

However, amongst those podcasts that are seasonal one can imagine a scenario whereby it could be desirable to present a seasonal show differently to other shows, to recommend or notify a user when a new season starts, or simply to recommend seasonal shows to users who typically like them.

\subsection{Is the seasonal nature of shows accurate?}

Again, taking examples of popular podcast shows, That Peter Crouch Podcast is a podcast discussing life as a British Premier League Footballer (amongst other things, \#passthepod)\cite{ThatPeterCrouch}. One of the episode titles is \emph{That Series 4 Announcement} - it's very clearly a seasonal show. And yet, according to the season tag in the RSS metadata, it only has one season.

TMSoft's White Noise Sleep Sounds is pretty much as you'd expect - a podcast with different white noise sleep sounds\cite{Tmsoft}. Not the kind of show you'd expect to be seasonal. However, according to the RSS feed it has three different seasons across over 150 episodes.

Perhaps the simplest way of classifying the likely size of the error in the seasonal field is to look at the episode title for evidence of seasonality. Episode titles such as S3E7 and Series 4 Announcement indicate that the show is likely to be seasonal. Looking across the entire Spotify podcast catalogue we find that around 1.5\% of podcast shows that aren't using the metadata field for indicating seasonality are naming their episodes as though they are seasonal.

Of course, there are other approaches to identifying seasonality without using the season metadata tag. By understanding the typical release cadence of a show, and then looking for unusually long breaks in publishing, we might be able to conclude that a podcast show is having a seasonal break. It is also possible to use episode transcripts to try to pick up references to the close of a season, or the start of a new one.

In our experience, at least some of these approaches are necessary in order to adequately reflect seasonality in podcasts. Effectively, we're advising against relying solely on the season metadata in the RSS feed.

\section{Category}

In 2019, Apple announced that it was updating its podcast categories - adding new categories such as True Crime, Fiction \& History as well as modifying and removing some existing categories\cite{CategoryChange}. The category of podcast is set by the creator and is widely used to allow users to browse podcast shows that might be of interest to them. It could obviously also be used to enhance recommendations. Example top-level podcast categories include Business, Comedy, Education and Science\cite{CategoryList}. Example sub-level podcast categories include Wrestling, Music Commentary and Astronomy\cite{CategoryList}.
 
There are a couple of complicating factors when considering the category assigned to a specific podcast. Firstly, as shown in figure 3, each podcast can have multiple top-level category labels.

\begin{figure}[h]
  \centering
  \includegraphics[width=\linewidth]{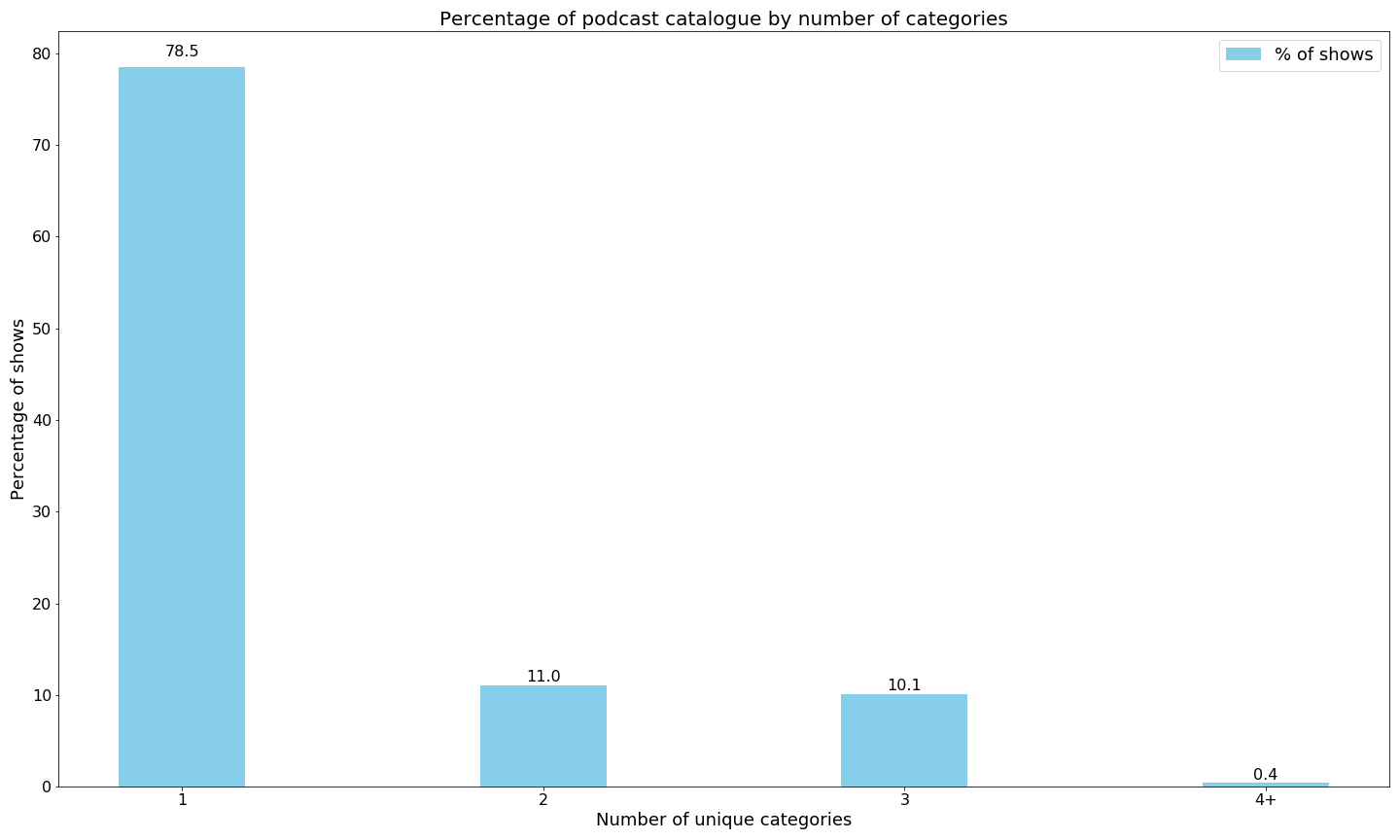}
  \caption{Number of categories per show}
  \Description{Podcast shows tend to have fewer than 3 categories}
\end{figure}

Given the range of number of categories that an individual podcast has, we have to be careful in our treatment of category as a recommendation vector. It would be easy to be more likely to recommend a podcast that appears in multiple categories, and hence encourage creators to list their podcast as belonging to multiple categories that don't especially fit. It's possible that we're already seeing podcasts label themselves as the union of all of the categories that they could fall under, rather than the intersection.

Additionally, one of the problems with the existing top-level podcast categories is how broad they really are. A popular category is Kids \& Family. Popular podcasts in this category include sleep meditation, celebrity interviews with a celebrity and their mum, parenting advice podcasts and children's stories.

The category Religion \& Spirituality includes sermons, celebrities recounting the best advice they've ever received, histories of the occult and discussions on American politics.

With categories this broad, it's possible that the range of topics covered within each one might mean they don't provide great input for the user. User research has led us to the following test for determining whether a category is likely to be useful or whether it needs modifying: could anybody reasonably say "I'm a fan of/into X", where X is the category name. It'd be fairly unusual and a little weird for somebody to say "I'm a fan of leisure". But somebody might say "I'm really into true crime". As such, we believe that True Crime is a much better category than Leisure. Generally, most of the sub-categories pass this test (much more so than the top-level categories).

In short, we advise caution when using the category label for any user-facing purpose.

\section{One Show per RSS Feed}

My Favorite Murder with Karen Kilgariff and Georgia Hardstark is a very popular True Crime podcast\cite{MFM}. 

\begin{figure}[h]
  \centering
  \includegraphics[width=\linewidth]{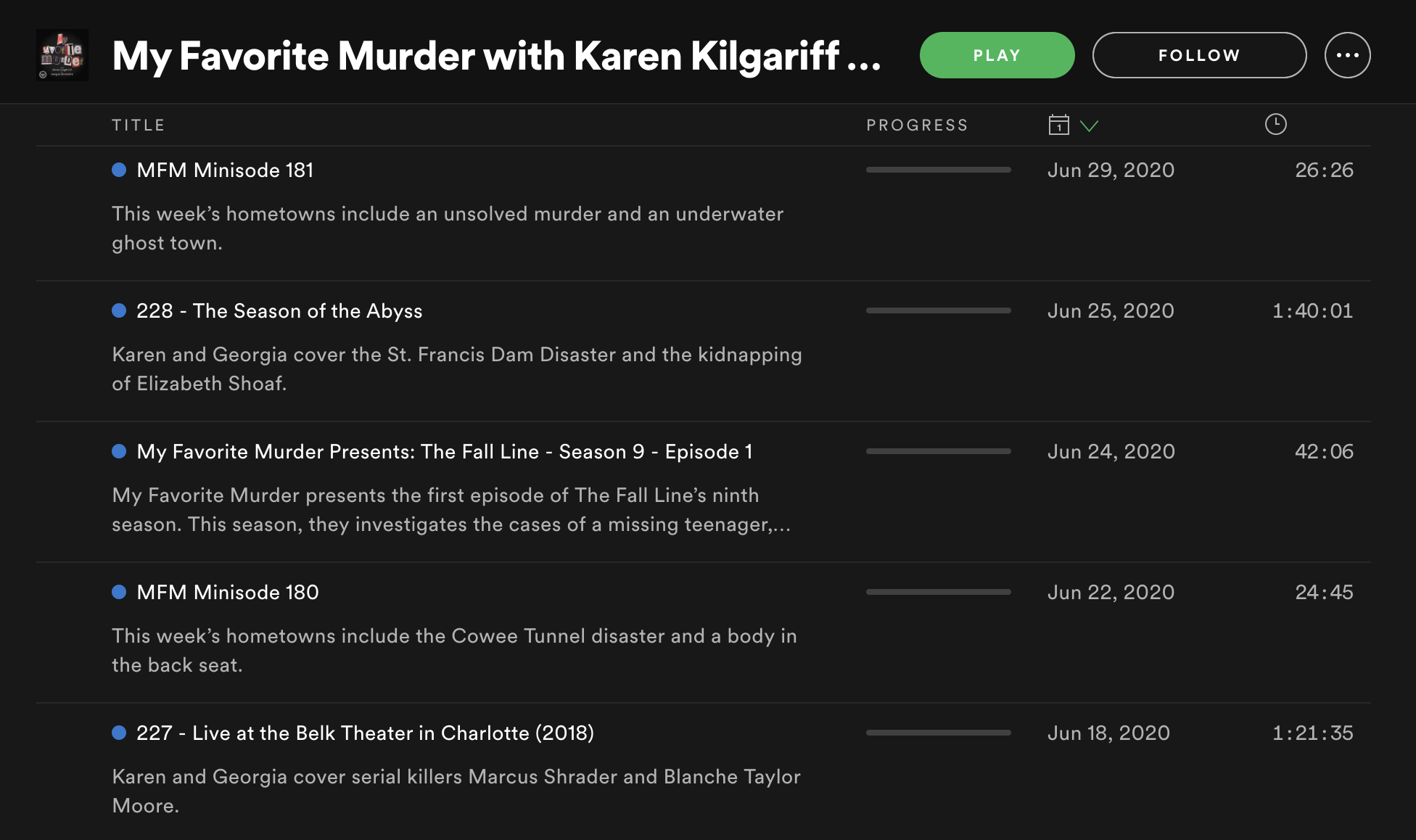}
  \caption{Screenshot of My Favorite Murder with Karen Kilgariff and Georgia Hardstark}
  \Description{There are four different episode types here, the main episode, the minisode, the live show and a feed drop.}
\end{figure}

As you can see in figure 4, there are at least four distinct episode types contained in the My Favorite Murder feed.

\begin{itemize}
    \item{\verb|Main episodes|}: In the figure, \emph{228 - The Season of the Abyss}. Are published every week and are generally between 90 and 120 minutes long and cover a couple of famous crimes/murders.
    \item{\verb|Minisodes|}: In the figure, \emph{MFM Minisode 181}. Are published every week and are generally less than 30 minutes long. Were introduced sporadically early on in the show's history, but have since become a staple part of the regular feed.
    \item{\verb|Live shows|}: In the figure, \emph{Live at the Belk Theater in Charlotte (2018)}. Are published roughly every month and are generally around as long as the main episodes. They follow the same format as the main episodes.    
    \item{\verb|Feed drops|}: In the figure, \emph{My Favorite Murder Presents: The Fall Line - Season 9 - Episode 1}. Are published rarely - a promotional device used by podcasts. Including other podcast shows in their feed allows a podcast to introduce their existing audience to another show.
\end{itemize}

If we assumed that all of these episodes belonged to the same entity and treated them as such in the client and in our recommendations algorithms, we may get confused and end up recommending a feed drop or a live show to a user who might have little interest in these. Equally, we could confuse ourselves into wondering why the episode lengths are so disparate, without understanding the nuance of the multiple episode types.

It's possible to think of each of these different episode types as belonging to a separate sub-show - a subdivision of the show that falls above episodes but below the show in the hierarchy of a podcast. This kind of behaviour is much more common on bigger podcast shows (introducing additional sub-shows as the show gains traction) than on less popular or newer shows.

Imagining these sub-shows as their own show-like entities could lead to interesting strategies for podcast recommendations and client design, such as recommending episodes only from a specific sub-show type.

\section{Conclusion}

There is a wealth of metadata available when building user experiences for podcasts. The season number, the category, and the consumption order are all fields that might seem useful when building recommendations, or any other user-facing feature. However, the metadata fields associated with a given podcast are creator-inputted data and should be treated as such when building anything using them.

\section{Citations and Bibliographies}

\begin{acks}
Thank you to Ching-Wei Chen, Maria Stone, Greg Herman, Bill Irwin, Ashley Skov, Laura Pezzini and Matt Lieber of Spotify for reviewing the work.
Thank you to Filipe La Ruina for providing and explaining data sets.
And thank you to Rachael and Arthur. Just generally, for many things.
\end{acks}

\bibliographystyle{ACM-Reference-Format}
\bibliography{sample-base}

\appendix
\section{Example RSS Feed\cite{RSSExample}}
\subsection{Example Episode}
\begin{verbatim}
    <item>
      <itunes:episodeType>full</itunes:episodeType>
      <itunes:episode>4</itunes:episode>
      <itunes:season>2</itunes:season>
      <title>S02 EP04 Mt. Hood, Oregon</title>
      <description>
        Tips for trekking around the tallest mountain in Oregon
      </description>
      <enclosure
        length="8727310" 
        type="audio/x-m4a" 
        url="http://example.com/podcasts/everything/mthood.m4a"
      />
      <guid>aae20190606</guid>
      <pubDate>Tue, 07 May 2019 12:00:00 GMT</pubDate>
      <itunes:duration>1024</itunes:duration>
      <itunes:explicit>false</itunes:explicit>
    </item>    
\end{verbatim}
\subsection{Example show metadata}
\begin{verbatim}
    <?xml version="1.0" encoding="UTF-8"?>
<rss version="2.0" xmlns:itunes="http://www.itunes.com/dtds/podcast-1.0.dtd" xmlns:content="http://purl.org/rss/1.0/modules/content/">
  <channel>
    <title>Hiking Treks</title>
    <link>https://www.apple.com/itunes/podcasts/</link>
    <language>en-us</language>
    <copyright>&#169; 2020 John Appleseed</copyright>
    <itunes:author>The Sunset Explorers</itunes:author>
    <description>
      Love to get outdoors and discover nature&apos;s treasures? Hiking Treks is the
      show for you. We review hikes and excursions, review outdoor gear and interview
      a variety of naturalists and adventurers. Look for new episodes each week.
    </description>
    <itunes:type>serial</itunes:type>
    <itunes:owner>
      <itunes:name>Sunset Explorers</itunes:name>
      <itunes:email>mountainscape@icloud.com</itunes:email>
    </itunes:owner>
    <itunes:image
      href="https://applehosted.podcasts.apple.com/hiking_treks/artwork.png"
    />
    <itunes:category text="Sports">
      <itunes:category text="Wilderness"/>
    </itunes:category>
    <itunes:explicit>false</itunes:explicit>
    <!-- One item per episode -->
    <!-- Item example given in the episode above -->
    <item>....</item>
  </channel>
</rss>
\end{verbatim}
\end{document}